\documentstyle[12pt]{article}

\newcommand{\be}{\begin{equation}}
\newcommand{\ee}{\end{equation}}
\newcommand{\bea}{\begin{eqnarray}}
\newcommand{\eea}{\end{eqnarray}}
\newcommand{\nn}{\nonumber \\}

\newcommand{\ba}{\begin{array}}
\newcommand{\ea}{\end{array}}
\newcommand{\vs}[1]{\vspace{#1 mm}}

\renewcommand{\a}{\alpha}

\def\bbox{{\,\lower0.9pt\vbox{\hrule \hbox{\vrule height 0.2 cm
\hskip 0.2 cm \vrule height 0.2 cm}\hrule}\,}}
\newcommand{\dsl}{\pa \kern-0.5em /}

\newcommand{\pa}{\partial}

\font\mybb=msbm10 at 12pt
\def\bb#1{\hbox{\mybb#1}}

\def\bR {\bb{R}}
\def\bE {\bb{E}}

\def\a{\alpha}

\def\a{\alpha}

\def\r{\rho}                                     

\def\car{{\cal R}}

\begin{document}

\topmargin 0pt
\oddsidemargin 5mm

\renewcommand{\thefootnote}{\fnsymbol{footnote}}
\begin{titlepage}

\setcounter{page}{0}
\begin{flushright}
{\sc KUL-TF}-98/30\\
hep-th/9807137\\
Revised, Nov. 98.
\end{flushright}

\vs{5}
\begin{center}
{\Large The domain-wall/QFT correspondence}
\vs{10}

{\large
H.J. Boonstra$^1$\footnote{harm.boonstra@fys.kuleuven.ac.be}
, K. Skenderis$^1$\footnote{kostas.skenderis@fys.kuleuven.ac.be}
 and  P.K. Townsend$^2$\footnote{
pkt10@damtp.cam.ac.uk, 
on leave from DAMTP,  University of Cambridge, U.K.} } \\
\vs{5}
${}^1${\em Instituut voor Theoretische Fysica\\
           KU Leuven, Celestijnenlaan 200D,\\
           3001 Heverlee, Belgium}\\
${}^2${\em Institute for Theoretical Physics,\\
University of California at Santa Barbara,\\ 
CA 93106, USA.}
\end{center}
\vs{10}
\centerline{{\bf Abstract}}

We extend the correspondence between adS-supergravities and superconformal
field theories on the adS boundary to a correspondence between gauged
supergravities (typically with non-compact gauge groups) and quantum field
theories on domain walls. 

\end{titlepage}
\newpage
\renewcommand{\thefootnote}{\arabic{footnote}}
\setcounter{footnote}{0}

\section{Introduction}

Evidence is currently accumulating for a conjectured equivalence between 
M-theory or IIB superstring theory in an anti-de Sitter (adS) background and
a superconformal field theory (SCFT) at the adS boundary \cite{malda}
(for related earlier work see \cite{kleb}, \cite{KSKS}). The 
isometry group of the KK vacuum acts as the superconformal group on the SCFT at
the adS boundary in the manner envisaged in earlier studies of singleton field
theories \cite{fronsdal} and branes `at the end of the universe' \cite{bdps}.
In the new approach the SCFT describes the dynamics of $N$ near-coincident
branes in the low-energy limit (or equivalently in the limit of decoupling
gravity). This limit corresponds to the
near-horizon limit of the corresponding brane solution of D=11 or IIB
supergravity, which turns out to be one of the well-known Kaluza-Klein (KK)
compactifications to an adS spacetime \cite{GT}. The (conjectured) equivalence
of the bulk supergravity theory (more precisely the underlying  M-theory or
superstring theory) to the SCFT boundary theory nicely illustrates the
holography principle\footnote{The equivalence between a KK supergravity and a 
SCFT at the adS boundary was actually already conjectured in \cite{bdps} but it
was then viewed not as an illustration of holography but as a generalization to
branes of the fact that the conformal field theory on a string worldsheet
encodes D=10 supergravity scattering amplitudes.} that is widely believed to be
a feature of any consistent theory of quantum gravity \cite{holography}.
However, this principle also suggests that the adS/CFT correspondence is just a
special case of a more general correspondence between supergravity theories (at
least those that are effective theories for some consistent quantum theory) and
quantum field theories in one lower dimension. 

One clue to a possible generalization of the adS/CFT correspondence is the 
fact that the adS metric in horospherical coordinates is a special case of
a domain wall metric \cite{lpt}. The isometry group of the generic 
D-dimensional
domain wall spacetime is the Poincar{\'e} group in $(D-1)$ dimensions. In the
supergravity context a domain wall typically preserves half the supersymmetry
and hence admits a super-Poincar{\' e} isometry group. If the supergravity 
theory arises via KK compactification on some space $B$ then the
KK domain wall `vacuum' is additionally invariant under the isometries
of $B$. The adS case is special in that the KK vacuum is invariant under
a larger $adS_D$ supergroup that contains as a proper subgroup the product 
of the $(D-1)$-dimensional Poincar{\'e} group with the isometry group of $B$.
This precisely corresponds to the fact that a $(D-1)$-dimensional
superconformal field theory is a special case of a $(D-1)$-dimensional
supersymmetric quantum field theory (QFT), for which the invariance
group is generically just the product of the $(D-1)$-dimensional 
super-Poincar{\'e} group with its R-symmetry group. The latter is naturally
identified with the isometry group of $B$ which is also, according to
Kaluza-Klein theory, the gauge group of the $D$-dimensional 
supergravity theory admitting the domain wall solution. 

We are thus led to investigate whether there are KK compactifications of
D=10 or D=11 supergravity theories to domain wall `vacua' analogous to 
compactifications to an adS spacetime. One cannot expect such a solution to 
preserve all supersymmetries but experience with branes suggests that
it might preserve 1/2 supersymmetry. In fact, a number of examples of this
type are already known and, in close analogy to the adS case, 
they can be interpreted as the `near-horizon' limits of brane solutions of 
D=11 or D=10 supergravity theories. For example, the `near-horizon' limit of 
the
NS-5-brane of N=1 supergravity yields an $S^3$ compactification of N=1 D=10
supergravity to a D=7 spacetime that can be interpreted as a  domain-wall
solution of the effective D=7 theory \cite{GT,cow, boonstra}. The latter can be
identified with the $SU(2)$-gauged D=7 supergravity \cite{vanN} coupled to an
$SU(2)$  super-Yang-Mills (SYM) multiplet, in accord with the fact that the 
isometry
group of $S^3$ is $SU(2)\times SU(2)$ (the same $S^3$ compactification is also
applicable to the NS-5-branes of Type II supergravities but we postpone 
discussion of these cases). A new example that we shall discuss here is an 
$S^2$ compactification of IIA supergravity to a D=8 domain wall spacetime. 
It can be found as a `near-horizon' limit of the IIA D6-brane solution. 
The effective D=8 supergravity theory is the $SU(2)$-gauged maximal 
supergravity\footnote{This was originally obtained as an $S^3$ reduction 
of D=11
supergravity \cite{SS}, but it can be viewed as an $S^2$ compactification 
of IIA
supergravity in which the RR 2-form field strength is proportional to the 
volume
of the 2-sphere.}. As adS supergroups exist only for  $D\le7$ it follows that
this, or any other, D=8 supergravity theory  cannot admit a supersymmetric adS
vacuum (by which we mean one for which the isometry group is an adS
supergroup). However, as we shall show, it does admit a 1/2
supersymmetric domain wall vacuum, and this domain  wall solution is precisely
the one found in the `near-horizon' limit of the D6-brane.

The term `near-horizon' is placed in quotes in the above paragraph 
because what is essential for the argument is not the existence of a 
horizon but rather of a second asymptotic region near the core of the 
brane into which spacelike geodesics can be continued indefinitely.
This is a feature of the NS-5-brane in string-frame because in this frame
the singularity at the core is pushed out to infinity. It is not a 
feature of the D6-brane in either the string frame or the Einstein frame
but there {\it is} a frame, the `dual' frame, in which the singularity is again
pushed out to infinity, and in this frame one finds a `vacuum' solution of IIA 
supergravity with an $adS_8\times S^2$ 10-metric \cite{boonstra, boonstrab}.
The `dual' frame can be defined for general $p$ as the one for
which the tension of the dual $(6-p)$-brane is independent of the 
dilaton. This implies that in the effective action the function of the dilaton 
multiplying the Einstein term is the same as the function  multiplying the 
dual of the $(p+2)$-form field strength. It is a general property of p-brane 
solutions, except when $p=5$, that the dual frame metric is the product of an 
adS space with a sphere \cite{DGT,boonstra,boonstrab}. 
When $p=5$ one finds that the 
adS space is replaced by a Minkowski spacetime, so this case requires a 
separate discussion. In all cases, however, there is an `internal' infinity in
the dual metric, and hence a second asymptotic region near the p-brane core. 
The importance of this is that the effective supergravity in this asymptotic 
region must, to the extent to which the supergravity approximation remains 
valid, describe the `internal' dynamics on the brane\footnote{As against the 
`external' dynamics of the brane's motion in the D=10 or D=11 spacetime, 
which  is described by the Born-Infeld type brane actions for the Goldstone 
modes of broken supertranslations.}. 

One thus arrives at the
(tentative) conclusion that the QFT describing the `internal' dynamics of $N$
coincident branes is equivalent to the supergravity theory in the 
`near-horizon' region of the corresponding supergravity brane solution
(but in a given region of the parameter space there is only one 
weakly coupled theory).
This is essentially the argument of \cite{malda} in support of the adS/CFT
correspondence, for which an extension to non-conformal cases was considered 
in \cite{maldatwo}. One purpose of this paper is to show how 
the dual frame allows a uniform discussion
of both the conformal and non-conformal cases. The conformal cases are those
for which the dual frame is self-dual. In all cases an essential ingredient
of the correspondence between the supergravity theory and the QFT is that
the conserved currents of the QFT couple to the bulk supergravity fields.
One therefore needs to specify some embedding of the brane worldvolume in
the bulk spacetime. In other words, where do we put the branes?

In the adS/CFT correspondence, it is natural to place the CFT at the adS 
boundary because this hypersurface is a fixed surface under the action
of the conformal group on the adS spacetime. The only other hypersurface 
with this property is the horizon. However, one can define a Minkowski space 
QFT with non-linearly realized conformal invariance on any horosphere
\cite{Kal}. The $adS_{p+2}$ metric in horospherical coordinates is
\be
ds^2 = u^2 ds^2(\bE^{(p,1)}) + u^{-2}du^2
\ee
and the horospheres are the hypersurfaces of constant $u$. A typical horosphere
is shown on the Carter-Penrose diagram of $adS_{p+2}$ in the figure below. 
\begin{figure}[h]
\begin{center}
\setlength{\unitlength}{1cm}
\begin{picture}(5,10)
 
\put(1,9.5){\line(1,-1){3}}
\put(1,3.5){\line(1,1){3}}
\put(1,3.5){\line(1,-1){3}}
 
\put(1,9.5){\line(1,1){0.5}}
\put(4,0.5){\line(-1,-1){0.5}}
 
\put(1,0){\line(0,1){10}}
\put(4,0){\line(0,1){10}}
 
\qbezier(4,0.5)(2.7,3.5)(4,6.5)

\end{picture}
\setlength{\unitlength}{1pt}
\caption{{\it Carter-Penrose diagram of anti-de Sitter spacetime. The
diagonal lines are Killing horizons. The curved line is a horosphere.}}
\end{center}
\end{figure}
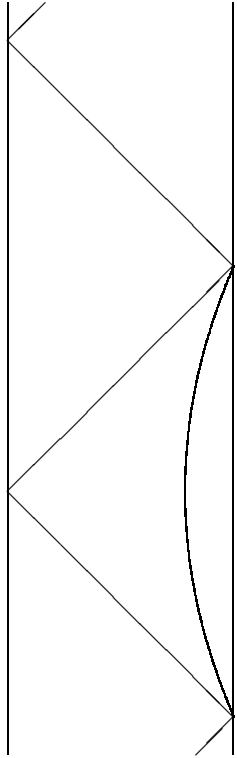
\noindent
The horospheres form a one-parameter class of Minkowski hypersurfaces
interpolating between the Killing horizon and the adS boundary, the parameter
being the radial adS coordinate. A scale transformation shifts the value of
this radius, i.e. it takes a QFT in one vacuum to a QFT in  another vacuum.
In the IR or UV limit one is driven to the adS horizon or its
boundary, respectively, these  being the two fixed points in the
space of vacua. Almost the same  picture carries over to the domain-wall/QFT
correspondence with the  difference that the choice of horosphere for the
Minkowski vacuum now corresponds not to a choice of the  vacuum of a QFT with
non-linearly realized conformal symmetry but rather to a choice of coupling
constant of a non-conformal QFT. Thus, in the non-conformal case the
interpolation between the adS Killing horizon and its boundary corresponds to
an interpolation between strong and weak coupling (or vice-versa, depending on
the case). Since the coupling constant is dimensionful this is again 
equivalent to a choice of scale.  

In the adS/CFT correspondence, operators of the CFT couple to fields
of the KK supergravity about the adS background, and correlation functions of
these operators are related to classical solutions of the KK supergravity
equations with boundary conditions specified at the adS boundary. 
We expect the same to be true in the domain-wall/QFT
correspondence but the graviton in this background no longer belongs to the
graviton supermultiplet of an adS-supergravity. Rather, it must belong to the
graviton supermultiplet of a class of gauged supergravity theories for which
the vacuum is instead a 1/2 supersymmetric domain wall spacetime. 
We further expect the QFT on the worldvolume of 
coincident Dp-branes to be encoded in a KK supergravity theory with a 1/2
supersymmetric domain wall vacuum and another purpose of this paper is to
identify the relevant lower-dimensional gauged supergravity theory in each 
case. We shall see that, in each case, the gauge group of the
supergravity theory coincides with the R-symmetry group of the
equivalent QFT.

\section{`Near-horizon' limit in the dual frame}

We shall be interested in a limit of (IIA, IIB, or N=1) superstring
theory for which the dynamics is well-approximated by an effective
supergravity action. The part of this action relevant to the Dp-brane contains
the metric, dilaton and a Ramond-Ramond field strength, which can be either
a $(p+2)$-form (in which case the Dp-brane solution is `electric') or a
$(8-p)$-form (in which case the Dp-brane solution is `magnetic'). For the
latter choice, and in the string frame, the action is
\be\label{actionten}
S = {1\over \a'{}^4} 
\int d^{10}x\, \sqrt{-g} \big[ e^{-2\phi}\big( R + 4(\partial \phi)^2 \big) - 
 {1\over 2 (8-p)!}\, |F_{8-p}|^2 \big]
\ee
The validity of the supergravity approximation requires that we take the limits
\be\label{limit}
\alpha'\rightarrow 0 \qquad g_s \rightarrow 0
\ee
where $\alpha'$ is the inverse string tension and $g_s= e^{\phi_\infty}$
is the string coupling constant (with $\phi_\infty$ the asymptotic
value of the dilaton field $\phi$). 

The equations of motion of the above action have the solution
\bea\label{pbrane}
ds^2_{st} &=& H^{-1/2} ds^2(\bE^{(p,1)}) + H^{1/2}ds^2(\bE^{(9-p)})]\nn
e^\phi &=& g_sH^{(3-p)/4} \nn
F_{8-p} &=& g_s^{-1} \star dH,
\eea
where $*$ is the Hodge dual of $\bE^{(9-p)}$ and $H$ is harmonic on
$\bE^{(9-p)}$. Let $r$ be the distance from the origin of $\bE^{(9-p)}$.
The choice
\be
H = 1 + g_sN\left(\sqrt{\a'}\over r\right)^{(7-p)}
\ee
then yields the long-range fields of $N$ infinite parallel planar Dp-branes
near the origin. 

We now consider the low energy limit\cite{malda, maldatwo},
$\a' \to 0$, keeping fixed the mass of stretched strings,
$U=r/\a'$ (so $r\to 0$), and all (other than $r$) coordinates 
that appear in (\ref{pbrane}). In addition, we hold fixed 
the 't Hooft coupling constant,
\be \label{th}
g_{YM}^2N={\rm fixed},
\ee
where
\be
g_{YM}^2 = g_s (\a')^{(p-3)/2}.
\ee
This means that we must have $g_sN\rightarrow 0$ for $p<3$ and 
$g_sN\rightarrow \infty$ for $p>3$. For $p<3$ this requirement is satisfied for
any $N$ because we take $g_s\rightarrow 0$. For $p>3$ it can be satisfied only
if we also take the limit $N\rightarrow \infty$. For $N$ finite, 
(\ref{th}) is equivalent to keeping energies finite
on the worldvolume while taking the low-energy limit.
The $N\to \infty$ limit is the 't Hooft limit\cite{hooft}.
In either case, we now have
\be
H= 1 + {g_{YM}^2N\over (\alpha')^2 U^{(7-p)}} \rightarrow
g_{YM}^2N(\alpha')^{-2} U^{(p-7)}.
\ee

The string metric of (\ref{pbrane}) is singular at $U=0$, so it is
not yet clear whether the above `near-horizon' limit yields a limiting
supergravity solution in the way that it does in the D3-brane case \cite{GT}. 
We can circumvent this problem by considering the `dual frame' metric
\be \label{Dmetric}
g_{dual} = (e^{\phi}N)^{{2 \over p-7}} g_{st}.
\ee
In this frame the action (\ref{actionten}) is
\be
S = {N^2\over \a'{}^4} 
\int d^{10}x\, \sqrt{-g} (N e^{\phi})^\gamma \big[ R + {4(p-1)(p-4)\over
 (7-p)^2} (\partial \phi)^2 - {1\over 2 (8-p)!}\, 
{1 \over N^2}\, |F_{8-p}|^2 \big]
\ee
where 
\be\label{gam}
\gamma = 2(p{-}3)/(7{-}p)\, .
\ee
Note that the dual field strength couples to the dilaton in the same way as the
metric; hence the terminology `dual frame'. For $p=3$ the dual frame
coincides with the Einstein frame up to a power of N. The Dp-brane 
dual frame metric is
\be
ds^2_{dual}= \left(g_sN\right)^{2/(p-7)}
\left[H^{(5-p)/(p-7)} ds^2(\bE^{(p,1)}) + 
H^{2/(7-p)}ds^2(\bE^{(9-p)})\right]
\ee
for which the singularity at $U=0$ is now just a coordinate
singularity. The full `near-horizon' solution is
\bea \label{nearh}
ds^2_{dual} &=& \a'[ (g^2_{YM} N)^{-1} U^{5-p} ds^2(\bE^{(p,1)}) + U^{-2} dU^2
+ d\Omega^2_{(8-p)}]\nn
e^{\phi} &=&  {1 \over N} [(g_{YM}^2 N) U^{p-3}]^{(7-p)/4} \nn
F_{8-p} &=& (7-p) N (\a')^{(7-p)/2} vol(S^{8-p}).
\eea
The near-horizon metric is $adS_{p+2}\times S^{8-p}$ when $p\ne5$. 
When $p=5$ it is $\bE^{(6,1)}\times S^3$, so we exclude this case in
what now follows. We will return to the $p=5$ case later.

It is important to notice that all factors of $\a'$ cancel out at the end:
in the string worldsheet action the overall $\a'$ in the metric cancels
against the $\a'$ in the string tension, and in the effective
supergravity action the factors of $\a'$ coming from 
the factors of $\a'$ in (\ref{nearh}) cancel against the $\a'$ in Newton's 
constant. This cancellation allows us to now set $\alpha'=1$ in what follows. 

The adS metric can be put in a standard form by the 
introduction of a new radial coordinate $u$ defined by
\be \label{newvar}
u^2 = \car^2 (g_{YM}^2 N)^{-1} U^{5-p} \qquad [\car=2/(5-p)]\, .
\ee
Since both $U$ and $g_{YM}^2 N$ remain finite in the near-horizon
limit, so too does $u$. We now have
\bea\label{compac}
ds^2_{dual} &=& \frac{u^2}{\car^2} ds^2(\bE^{(p,1)}) 
+ \car^2 \frac{du^2}{u^2}
+ d\Omega^2_{(8-p)}\nn
e^{\phi} &=& {1 \over N} 
(g^2_{YM} N)^{(7-p)/2(5-p)}
({u/\car})^{(p-7)(p-3)/2(p-5)} \nn
F_{8-p} &=& (7-p) N vol(S^{8-p}).
\eea
We see from this form of the metric that $\car$ is the adS radius of
curvature. The hypersurface $u=0$ 
is a non-singular Killing horizon while the boundary of the adS space 
is at $u=\infty$. 

There is an UV/IR connection between the bulk and the boundary
theory. Specifically, it was shown in \cite{PP} that the
holographic energy scale of the boundary QFT is
\be\label{holog}
E \sim  {U^{(5-p)/2} \over g_{YM} N^{1/2}}\, .
\ee
This energy-distance relation leads to a holographic result for 
the number of states of string theory in $adS$ space \cite{SW,PP}. From 
(\ref{newvar}) we see that it is equivalent to
\be \label{disen}
E\sim u\, .
\ee
In other words, the scale $u$ introduced by the requirement that the
$adS$ metric in the dual frame take the same form for all $p$
(including the conformal $p=3$ case) is the holographic energy scale
of the boundary QFT! 
Thus, one may argue that the dual-frame is the ``holographic'' 
frame describing supergravity probes (the string frame 
being associated with D-brane probes). 

The gauge theory description is
valid provided that the effective dimensionless YM coupling constant 
(at the energy scale of interest)
\be
g_{eff} = g_{YM}N^{1/2} E^{(p-3)/2}
\ee
is small. Using (\ref{disen})-(\ref{holog}) we see that this requires
\be\label{condita}
[g_{YM}^2 N U^{p-3}]^{(5-p)}<<1.
\ee

On the other hand, string perturbation is valid when the dilaton 
(\ref{compac}) is small. This is always true 
in the large $N$ limit, except near $u=0$ for $p<3$ and $p=6$
or $u=\infty$ for $p=4$. Depending on the rate with which 
we approach these points SYM or supergravity will be valid
in some region of the parameter space, but eventually
the string coupling grows large and we have to pass to the strong
string-coupling dual.

The validity of the supergravity solution requires
that the effective string tension times the 
characteristic spacetime length is large. The latter can be read-off 
from (\ref{compac}) and is of order 1. 
Therefore, we get the condition  
\be
T_{dual} = (Ne^\phi)^{2/(7-p)} >> 1,
\ee
which implies,
\be 
g_{YM}^2 N U^{p-3} >>1
\ee
for the validity of the supergravity description of the D-brane
dynamics. This is the same as the condition found in \cite{maldatwo,PP} by
requiring the curvatures in the string frame to remain small.

The above conditions and their implications have been 
discussed in detail  in \cite{maldatwo, PP}. We provide a brief 
summary here for
$p=0,1,2,4,6$, as this will be helpful for the discussion to follow of
the associated supergravity domain wall spacetimes. 
The $p=0,1,2$ cases are similar in that the region close to the 
boundary (corresponding to the UV of the SYM) is described in terms 
of perturbative SYM since the effective coupling constant 
is small there. The IR limit of the SYM theory, which corresponds
(in the dual frame) to the region near the horizon at $u=0$, is a strong
coupling limit because $g_{eff}$ is large there. The effective string
coupling constant $e^\phi$ is also large in this region, so the
horizon limit is the strong string coupling limit. This means that to 
resolve the singularity in $e^\phi$ at the horizon one must pass to
the strong coupling dual theory. For $p=0,2$ this is M-theory while for
$p=1$ it is the dual IIB theory. In the latter case, after S-duality 
the D-string supergravity solution becomes the fundamental string
solution. The dilaton is then small near $u=0$ but there is now a curvature
singularity there which is resolved in the DVV 
matrix string theory \cite{DVV}. 
In the $p=4$ case the region close to
the boundary again corresponds to the UV of the SYM theory but this
theory is now strongly coupled there. The $D=10$ supergravity
description is valid for sufficiently large $N$ at any given distance
from the boundary but for any $N$, no matter how large, the string
coupling blows up as the adS boundary is approached. The $D=10$
supergravity description therefore fails in the latter limit but a
description in terms of $D=11$ supergravity may be valid 
because the singularity of the dilaton at the boundary
is resolved by the interpretation of $adS_6$ with 
a `linear' dilaton as $adS_7$ \cite{DGT}. Near the adS horizon the
supergravity description fails but the SYM description is valid
because the effective gauge coupling is small.

The $p=6$ case is rather different because the supergravity description is
apparently valid in the same near-horizon region in which the
SYM theory is weakly coupled. This would seem to lead to a 
contradiction, but the issue of decoupling 
gravity in the case of $D6$ branes is rather intricate \cite{sen, seiberg}.

\section{Gauged supergravities and domain-walls}

The dual frame formulation is natural from the supergravity 
point of view because the factorization of the geometry 
leads to an immediate identification of the lower dimensional 
gauged supergravity that the graviton is part of.

{}From the solution (\ref{compac})
we see that there is an $S^{8-p}$ compactification of the
D=10 theory to an effective gauged $(p+2)$-dimensional supergravity with
action of the form 
\be
S = N^2 \int d^{p+2}x\, \sqrt{-g} (N e^{\phi})^\gamma \big[ R +
{4(p-1)(p-4)\over(7-p)^2}(\partial\phi)^2
+ {1 \over 2} (9-p) (7-p) \big]\, .
\ee
where $\gamma$ is the constant given in (\ref{gam}).
We also deduce that the field equations of this action admit an
$adS_{p+2}$ vacuum with `linear' dilaton (as explained in \cite{DGT}, the
dilaton can be invariantly characterized in terms of a conformal Killing
potential). 

On passing to the Einstein frame (for $p\neq0$) we find the action
\be\label{effE}
S = N^2 
\int d^{p+2}x\, \sqrt{-g} \big[ R -{1\over2}(\partial\phi)^2
+{1 \over 2} (9-p) (7-p) N^{b} e^{a\phi}\big]
\ee
where
\be
a={-\sqrt{2}(p-3)\over\sqrt{p(9-p)}}, \qquad b={4(p-3) \over p (p-7)}\, .
\ee
In this frame the $adS_{p+2}$ linear dilaton vacuum is equivalent to the domain
wall solution studied in \cite{LPSS0,LPSS,lpt}.
The parameter $\Delta$ used in \cite{lpt} to characterize various kinds
of domain wall solutions is in our case ($D=p+2$),
\be
\Delta\equiv a^2 - {2(D-1)\over D-2}={-4(7-p)\over 9-p}
\ee
with $\Delta_{\rm adS}={-2(D-1)\over D-2}$ the value corresponding
to the effective Lagrangian which admits anti-de Sitter spacetime
as a solution. As expected, $\Delta=\Delta_{\rm adS}$
only for $p=3$. For all other $p$ the domain wall
is in the category $\Delta_{\rm adS} < \Delta < 0$ \cite{lpt}.
We now turn to an application of the above analysis to Type II D-p-branes. The
values of $p$ to which the above analysis is applicable are $p=0,1,2,4,6$. We
exclude $p=3$ as the domain wall vacuum is then the well-known
supersymmetric adS vacuum. We shall consider the $p=5$ case later.

\subsection{p-branes, $p\ne5$}

{}From the $p=0$ near-horizon limit we see that there is an $S^8$
compactification of IIA supergravity to a D=2 $SO(9)$ gauged maximal 
supergravity theory. The R-symmetry group of the corresponding D=1 QFT
is expected to be the largest subgroup of $SO(16)$ with a 16-dimensional
spinor representation \cite{BGT}; this is precisely $SO(9)$. 
Since very little is known about gauged supergravity theories in
D=2 we turn now to $p=1$. For $p=1$ we see that there is an $S^7$
compactification of, for example, IIA supergravity to a D=3 $SO(8)$ gauged
maximal supergravity. This is obviously the $S^1$ reduction of the de
Wit-Nicolai $SO(8)$ gauged N=8 supergravity in D=4; as shown in \cite{lpt}, the
$adS_4$ vacuum of the latter will descend to a domain wall solution in
D=3. In the IIB case the interpretation of the $SO(8)$-gauged D=3 supergravity
as a reduction of the de Wit-Nicolai theory is not available, and it may
well be a different theory; gauged D=3 supergravities have not yet
been systematically explored. Furthermore, from the type I fundamental 
string solution we expect that there should be a truncation of  
maximal $SO(8)$-gauged D=3 supergravity to a half-maximal one with 
the same gauge group. In all of these cases the corresponding D=2 QFT
has the expected $SO(8)$ R-symmetry group (identifying the left and right
$SO(8)$ groups). 
 
We now turn to $p=2$. In this case we find an $S^6$ compactification of IIA
supergravity which we can also view as an $S^6\times S^1$ compactification of
D=11 supergravity\footnote{This is not included in a
previous classification \cite{CRM} of compactifying solutions of D=11
supergravity to D=4 because the D=4 spacetime is a domain wall spacetime rather
than adS.}. The effective gauged D=4 supergravity must be one of those found by
Hull as these have recently been shown to be complete
\cite{fre}. Since the isometry group of $S^6$ is $SO(7)$ we might
expect the gauge group to be $SO(7)$ (which is also
the R-symmetry group of the D2-brane). The obvious
candidate is the $ISO(7)$ gauged supergravity \cite{hull} because only
the $SO(7)$ subgroup can be linearly realized. Moreover, this theory has
a potential \cite{hullw} of the right form to admit a
1/2 supersymmetric domain wall solution. The correctness of this
identification follows from the observations \cite{hullwarner} that the
non-compact gauged N=8 supergravity theories can be obtained by
`compactification' of D=11 supergravity on hyperboloids of constant negative
curvature, and that the contracted versions, such as $ISO(7)$, correspond to a
limit in which the hyperboloid degenerates to an infinite cylinder. The
$ISO(7)$ theory thus corresponds to a `compactification' on the cylinder 
$S^6 \times \bE^1$, but we may replace this by $S^6\times S^1$. 
The reason that the near-horizon limit of the D2-brane differs from that of the
M2-brane (for which the effective D=4 theory is the $SO(8)$ gauged
de Wit-Nicolai theory) is that the M2-brane harmonic function is harmonic on
$\bE^8$ whereas the D2-brane harmonic function is harmonic on $\bE^7$. Further
dimensional reduction will lead to functions that are harmonic on $\bE^{8-k}$ 
and hence to new `near-horizon' limits, for which the effective D=4 theory is a
$CSO(8-k,k)$ gauged supergravity (in the notation of
\cite{hullb})\footnote{The linearly realized gauge group in this case
is $SO(8-k)$; one would therefore expect the (non-conformal)
interactions of the associated D=3 QFT to break $SO(8)$ to $SO(8-k)$.}.

Passing over $p=3$ we come to $p=4$. In this case we find an $S^4$
compactification of IIA supergravity. As mentioned earlier, this is
just the $S^4$ compactification of D=11 supergravity 
in disguise. This is consistent with the fact that the $SO(5)$ gauged maximal 
D=6 supergravity is just the reduction on $S^1$ of the $SO(5)$ gauged maximal 
D=7 supergravity \cite{cownew}. It is also consistent with the fact
that the R-symmetry group of the D4-brane is $SO(5)$.
 
Passing over $p=5$ we come to $p=6$. Despite the problematic features of the
correspondence in this case, consideration of the `near-horizon' limit
of the D6-brane dual-frame solution still allows us to deduce the existence of 
an $S^2$ compactification of IIA supergravity to an $SU(2)$ gauged D=8 
supergravity (note that $SU(2)$ is also the R-symmetry group of the 
D6-brane).  We shall now argue that this theory is the one found by
Salam and Sezgin \cite{SS} from a generalized Scherk-Schwarz 
reduction on an SU(2) group manifold of D=11 supergravity. 

In the $S^2$ compactification of IIA supergravity there is a 2-form field
strength proportional to the volume form on $S^2$. But this 2-form is also the 
field strength of the KK gauge field arising in the $S^1$ compactification of
D=11 supergravity. It follows that the $S^2$ compactification of IIA
supergravity is equivalent to a compactification of D=11 supergravity on a
$U(1)$ bundle over $S^2$. For unit charge this is just the Hopf fibration of
$S^3$.  As confirmation of the equivalence of the two compactifications we now
observe that the gravity/dilaton sector of the effective gauged D=8 
supergravity
arising from an $S^2$ compactification of IIA supergravity is, according to
(\ref{effE}),
\be\label{d8a}
S=\int d^8 x\, \sqrt{-g} \big[ R -{1\over2}(\partial\phi)^2
+{3\over2} N^{-2} e^{-\phi}\big]\,.
\ee
This has exactly the same dilaton potential as the Salam-Sezgin action
if all the other scalar fields appearing in the latter are set to zero,
and if one identifies the SU(2) coupling constant $g$ with $2 N^{-2}$.

The domain wall solution of the field equations of (\ref{d8a}) is
\bea
ds^2 &=& e^{{1\over2}\rho} ds^2(\bE^{(6,1)}) + N^{-2}
e^{{3\over2}\rho}d\rho^2\nn
\phi &=& {3\over2}\rho\, ,
\label{D6dw}
\eea
where we use a coordinate $\rho$ for which the dilaton is linear.
This is the domain wall solution of \cite{lpt} for $D=8$
and $\Delta=-{4\over3}$. It is easily shown to be a 1/2 supersymmetric
solution of the $SU(2)$ gauged D=8 supergravity by using the supersymmetry
transformation rules given in \cite{SS}. On the other hand, the same solution 
can be found from the near-horizon limit (\ref{nearh}) of the D6 brane
solution by splitting off the (compact) internal part  and
passing to the Einstein frame. 

\subsection{Fivebranes}
 
Returning to (\ref{nearh}) for $p=5$, defining
\be 
\rho = \log \left(g_{YM}N^{1/2}\, U\right) \, ,
\ee
and rescaling the worldvolume coordinates by $g_{YM} N^{1/2}$
we have the `near-horizon' solution 
\bea
ds^2_{dual} &=& \a'[ds^2(\bE^{(5,1)}) + d\rho^2 + d\Omega^2_3] \nn
\phi &=& \rho - \log N \nn
{}F_3 &=& 2 N \a'\, vol(S^3)
\eea
This is $\bE^{(6,1)}\times S^3$ with a linear 
dilaton. Minkowski space does not have a boundary, so 
the issue of holography for fivebranes is more intricate
\cite{maldatwo,ABKS,PP}. Nevertheless, we deduce, as in the previous cases, 
that there is an $S^3$ compactification to D=7 \cite{GT}. 
In the context of N=1 D=10 supergravity the resulting D=7 supergravity theory 
has been identified as the $SU(2)$ gauged theory of \cite{vanN}, coupled
to an $SU(2)$ SYM theory, with the D=7 vacuum being the domain wall 
solution found in \cite{LPSS}. 

The same $S^3$ compactification is a solution of IIA and IIB supergravity,
now arising as the `near-horizon' limit of the IIA or IIB NS-5-brane. Let us
first consider the IIA case, which can be viewed as an $S^3\times S^1$
compactification of D=11 supergravity. From our analysis of the $S^6\times
S^1$ compactification in the previous section we would expect the effective
D=7 theory to be an $ISO(4)$ gauged D=7 maximal supergravity. No such theory
is currently known but there is an $SO(4,1)$ gauged D=7 maximal supergravity
\cite{PPVW}. There is also an $SO(3,2)$ theory. The known list cannot be
complete, however, because it does not include the $SU(2)$ gauged theory found
by $S^1$ reduction of the $SU(2)$ gauged D=8 theory; the former is probably a
contraction of the $SO(3,2)$ D=7 theory. It therefore seems possible that
there is an $ISO(4)$ theory awaiting construction, and that this theory is
the effective D=7 theory for the KK compactification of IIA supergravity
provided by the near-horizon limit of the NS-5-brane\footnote{On the
other hand, the R-symmetry group of the IIA 5-brane is $SO(5)$, which
is larger than the linearly realized $SO(4)$ subgroup of $ISO(4)$.}.  

In the IIB case the effective D=7 theory may well be a different 
(as yet unknown)
$SO(4)$-gauged D=7 maximal supergravity theory.  Note that the $SO(4)$ gauge
group is expected not only from the fact that this is the isometry  group of
$S^3$ but also from the fact that the R-symmetry group of the IIB NS-5-brane is
$SO(4)$. Of course the same analysis applies to the D5-brane, in which case
it is clear that the QFT in the domain-wall/QFT correspondence should be
a D=6 $SU(N)$ SYM theory on a hypersurface of constant 
$\phi$ in the domain-wall spacetime. 
As $\rho$ varies from $-\infty$ to $\infty$ the effective coupling constant 
varies from zero (weak coupling) to infinity (strong coupling). In contrast
to the $p\ne5$ cases discussed previously, there is now {\sl no asymmetry 
between  the weak and strong coupling limits}. Unlike the $p\ne5$ cases, there
is now an $\r \rightarrow -\rho$ isometry of the near-horizon 
spacetime.

As a final observation concerning fivebranes we note that there is a
generalization of the $S^3$ compactification of D=10 supergravity theories 
to an
$S^3\times S^3$ compactification which can be interpreted \cite{cow} as the
near-horizon limit of intersecting 5-branes. In the context of N=1 supergravity
(or the heterotic string theory) the $S^3\times S^3$ compactification yields
the $SO(4)$ gauged Freedman-Schwarz model \cite{ant,cham}. The same solution 
can
be used to compactify either IIA or IIB supergravity, and one may then wonder
what the effective D=4 theory is in these cases. The only obvious candidate is
the $SO(4,4)$ gauged D=4 maximal supergravity of \cite{hullb}. Unlike the 
$SO(8)$ theory, it can be truncated to the FS model. The $SO(4,4)$ theory has a
(non-supersymmetric) de Sitter vacuum that has been interpreted as a
`compactification' of D=11 supergravity on a hyperboloid \cite{hullwarner}, but
this does not preclude the possibility of other solutions of the same
theory having quite different higher-dimensional interpretations.

\section{Comments}

We have proposed that superstring or M-theory in certain domain wall spacetimes
is equivalent to a quantum field theory describing the internal dynamics on
$N$ coincident branes. The near-horizon limit of the
corresponding brane supergravity solution yields a compactification to the 
domain wall spacetime of the proposed equivalence. This proposal extends the 
usual adS/CFT correspondence by viewing the adS spacetime as a special case 
of a domain wall spacetime.

In the course of formulating this proposal we have added to the list
of gauged supergravity theories that can be interpreted as `near-horizon' 
limits of brane, or intersecting brane solutions of M-theory or 
superstring theories (in this paper we considered single brane 
solutions, but we expect that a similar analysis can be performed
for intersecting branes as well). 
Notably, we have found a role for the $SU(2)$-gauged
maximal D=8 supergravity, and the $ISO(7)$ gauged N=8 theory in D=4. As
observed earlier, the last case is just one in the $CSO(8-k,k)$ series
of gauged N=8 D=4 supergravity theories that can be found as
near-horizon limits of the $T^k$ reduction of the M2-brane. Toroidal reductions
of other branes can be expected to lead to similar series of non-compact
gaugings in other dimensions. For example, $T^k$ reductions of the M5-brane
and the IIB 3-brane are expected to be related to new $CSO(5-k,k)$ gauged D=7
supergravity theories and $CSO(6-k,k)$ gauged D=5 supergravities respectively
(the known non-compact gaugings of maximal D=5 supergravity can be found in
\cite{gun}). Several of these supergravities are expected to be
related to each other by dimensional reduction since the corresponding
brane configurations are related by T-dualities. 
As yet there is no known brane interpretation of the
non-contracted versions of these non-compact gauge groups.

{}One intriguing aspect of our results is the association of the dual 
frame metric with the holographic energy scale in the boundary QFT.
The fact that the dual brane has dilaton independent tension in this
frame indicates that elementary $(6{-}p)$ branes may play a role. 
This possibility was ruled out for toroidal compactifications 
in \cite{hull95} but spherical compactifications might lead to a
different conclusion. 

A prerequisite of any domain-wall/QFT correspondence is that the R-symmetry of
the supersymmetric QFT on the domain wall worldvolume match the gauge group 
of the equivalent gauged supergravity. In this paper we have seen evidence
this requirement is  met by all $p$-branes with $p\le6$, the
restriction on $p$ arising from the fact that for $p>7$ the harmonic functions
are not bounded at infinity. Nevertheless, the fact that branes with $p>7$ are
connected via dualities to $p<7$ branes suggests that a similar story should
hold for $p\ge7$. For example, the worldvolume field theory of a 7-brane
has an $SO(2)$ R-symmetry, which suggests the existence of an $SO(2)$-gauged
$D=9$ maximal supergravity admitting a 1/2 supersymmetric domain wall vacuum. 
The D=9 duality group is $Gl(2;\bR)$, which is just big enough to allow for an
$SO(2)$ gauging. The construction of this model will be described in a future 
publication \cite{BS}. The remaining two cases, $p=8$ and $p=9$, also fit the
pattern. In the $p=8$ case the domain wall is the D-8-brane of the massive
IIA supergravity. The R-symmetry group of the field theory on the domain wall
is trivial as is the gauge group of the bulk IIA supergravity. In the
$p=9$ case the domain wall is the `M-boundary' of the Ho\v{r}ava-Witten 
theory \cite{horwit}.
The R-symmetry group is again trivial, as is the gauge group of the bulk
D=11 supergravity theory. 

\section*{Acknowledgments}
KS would like to thank the organizers of the Amsterdam summer 
workshop on ``String theory and black holes'' for hospitality while
this work was being completed. We would 
like to thank Bas Peeters for collaboration during initial stages of this work.
PKT is grateful to N. Warner and KS to J. Barbon for helpful conversations.


\end{document}